\documentclass[%
 aip,
 amsmath,amssymb,
 reprint,
]{revtex4-1}

\usepackage{graphicx}
\usepackage{dcolumn}
\usepackage{bm}

\usepackage[utf8]{inputenc}
\usepackage[T1]{fontenc}
\usepackage{mathptmx}
\usepackage{etoolbox}

\makeatletter
\def\@email#1#2{%
 \endgroup
 \patchcmd{\titleblock@produce}
  {\frontmatter@RRAPformat}
  {\frontmatter@RRAPformat{\produce@RRAP{*#1\href{mailto:#2}{#2}}}\frontmatter@RRAPformat}
  {}{}
}%

\newcommand*{\citen}[1]{%
  \begingroup
    \romannumeral-`\x 
    \setcitestyle{numbers}%
    \cite{#1}%
  \endgroup   
}

\draft 

\begin{document}


\title[]{Determination of Local Short-Range Order in TiVNbHf(Al)}
\author{Michael Xu}
\author{Shaolou Wei}
\author{Cemal C. Tasan}
\author{James M. LeBeau}
\affiliation{ 
Department of Materials Science and Engineering, Massachusetts Institute of Technology, Cambridge, MA 02139}
\email[Corresponding author, Email: ]{\tt lebeau@mit.edu}


\date{\today}

\begin{abstract}

The presence of short-range chemical order can be a key factor in determining the mechanical behavior of metals, but directly and unambiguously determining its distribution in complex concentrated alloy systems can be challenging. Here, we directly identify and quantify chemical order in the globally single phase BCC-TiVNbHf(Al) system using aberration corrected scanning transmission electron microscopy (STEM) paired with spatial statistics methods. To overcome the difficulties of short-range order (SRO) quantification with STEM when the components of an alloy exhibit large atomic number differences and near equiatomic ratios, ``null hypothesis'' tests are used to separate experiment from a random chemical distribution. Experiment is found to deviate from both the case of an ideal random solid solution and a fully ordered structure with statistical significance. We also identify local chemical order in TiVNbHf and confirm and quantify the enhancement of SRO with the addition of Al. These results provide insights into local chemical order in the promising TiVNbHf(Al) refractory alloys while highlighting the utility of spatial statistics in characterizing nanoscale SRO in compositionally complex systems. 

\end{abstract}

\maketitle 

Refractory high entropy alloys (RHEAs), composed of multi-principal refractory components, have garnered significant interest due to their high-temperature performance and potential for hierarchical control of constituent phases and microstructures at a variety of length scales \cite{senkov2018reviewRHEASRO}. Notably, the presence and manipulation of ordered precipitates -- down to the nanoscale -- improves ductility and performance of several alloys in this design space \cite{senkov_brittle_BCCB2_2018, Soni2020TiNbTaAlVZr}. 
Of particular interest are RHEAs based on Ti, V, Nb, and Hf that have been shown to exhibit high strength and ductility at room temperature with the potential for local chemical order \cite{wu_tinbhfv_2014_ductilebcc, Wei_naturalmixing_2020, Bu_2021_fluctuations}. In these and similar alloys, the existence of local chemical fluctuations, either through ordering or clustering, have been predicted and connected to the favorable mechanical response seen in these systems \cite{Bu_2021_fluctuations, an2020spinodalbeta}. Such local fluctuations and associated increased yield strength are thought to arise from the varying constituent elements' chemical affinities and site preferences, specifically among Ti, Nb, and Hf \cite{Xun2023sroRHEA}. 

Control of the local order in Ti- and Nb-containing RHEAs has been enabled through the addition of Al, where properties including strength, ductility, density, and oxidation resistance have been maintained or improved \cite{2014_senkov_jom_al_containing, 2014_senkov_acta_al_effect, senkov2018reviewRHEASRO, wu2021predictionRHEAsro, Liu2020tivnbal, 2022_zhu_matlet_alb2RHEA, yurchenko2022aladdition, 2022_khan_ductile_al_rhea}. In an AlTiNbZrW system, for example, a microstructure consisting of B2 nanoprecipitates in a BCC matrix has been attributed to exceptional room temperature and high temperature specific yield strength and plasticity \cite{2022_khan_ductile_al_rhea}. Similarly, Al-modulation has been shown to provide a systematic increase in yield strength and elongation in an AlNbTiZr system due to dislocation interactions with SRO \cite{yurchenko2022aladdition}. Moreover, the elastic response of these systems has been suggested to depend on local thermodynamically stable configurations of order and disorder \cite{Qiu_Chen_Naihua_Zhou_Hu_Sun_2021}. While controlling the formation of order in refractory high entropy alloys provides a promising route for microstructural and property engineering, direct,nanoscale investigations of these phenomena are required to provide new insights and model validation.


While comparing the binary compound formation enthalpies and mixing enthalpies across Ti, V, Nb, and Hf suggests that local short-range order (SRO) should be possible ($\Delta H_{f}^{binary}\approx -5$ to $4$ kJ/mol and $\Delta H_{mix}\approx -2$ to $4$ kJ/mol) \cite{prxbinaryformationenthalpy, Takeuchi_Inoue_2005_enthalpy2}, this has not been directly confirmed to the best of our knowledge. Moreover, characterizing SRO in these systems using techniques including STEM, X-ray diffraction, and neutron diffraction remains a challenge in part due to the presence of extremely local chemical fluctuations as well as significant differences in atomic number \cite{Chen_2019_xraydifficulty}. To complicate matters, the stability of B2 SRO/LRO has been suggested to arise due to a strong sublattice site disorder from the multi-component nature of these alloys \cite{kormann2021predictionAlTiSRO}, raising the question of how to differentiate order from disorder with confidence. 



In this work, we show the presence of extremely local ($<$2 nm) SRO in the refractory high entropy Ti$_{38}$V$_{15}$Nb$_{23}$Hf$_{24}$ alloy using a combination of atomic resolution STEM imaging and spatial statistics methods \cite{Xu_Kumar_LeBeau_2022}. To do so, we apply the Moran's I statistic, a two-dimensional autocorrelation analogous to the Warren-Cowley parameter. By establishing upper and lower ``bounds'' on the measurements obtained from atomic resolution STEM, local random chemical fluctuations in a ideal disordered random solid solution (RSS) are shown to be inconsistent with experiment. Specifically, the non-random, sublattice-based order between (TiV) and (NbHf) is found to be consistent with predicted site occupancies and bond preferences \cite{kormann2021predictionAlTiSRO}. Finally, we show that the addition of Al leads to a small but statistically significant enhancement of short-range ordering. Overall, the results of this study demonstrate the power of STEM to determine and quantify SRO at extremely short length scales in chemically complex systems. 

\begin{figure}
\includegraphics{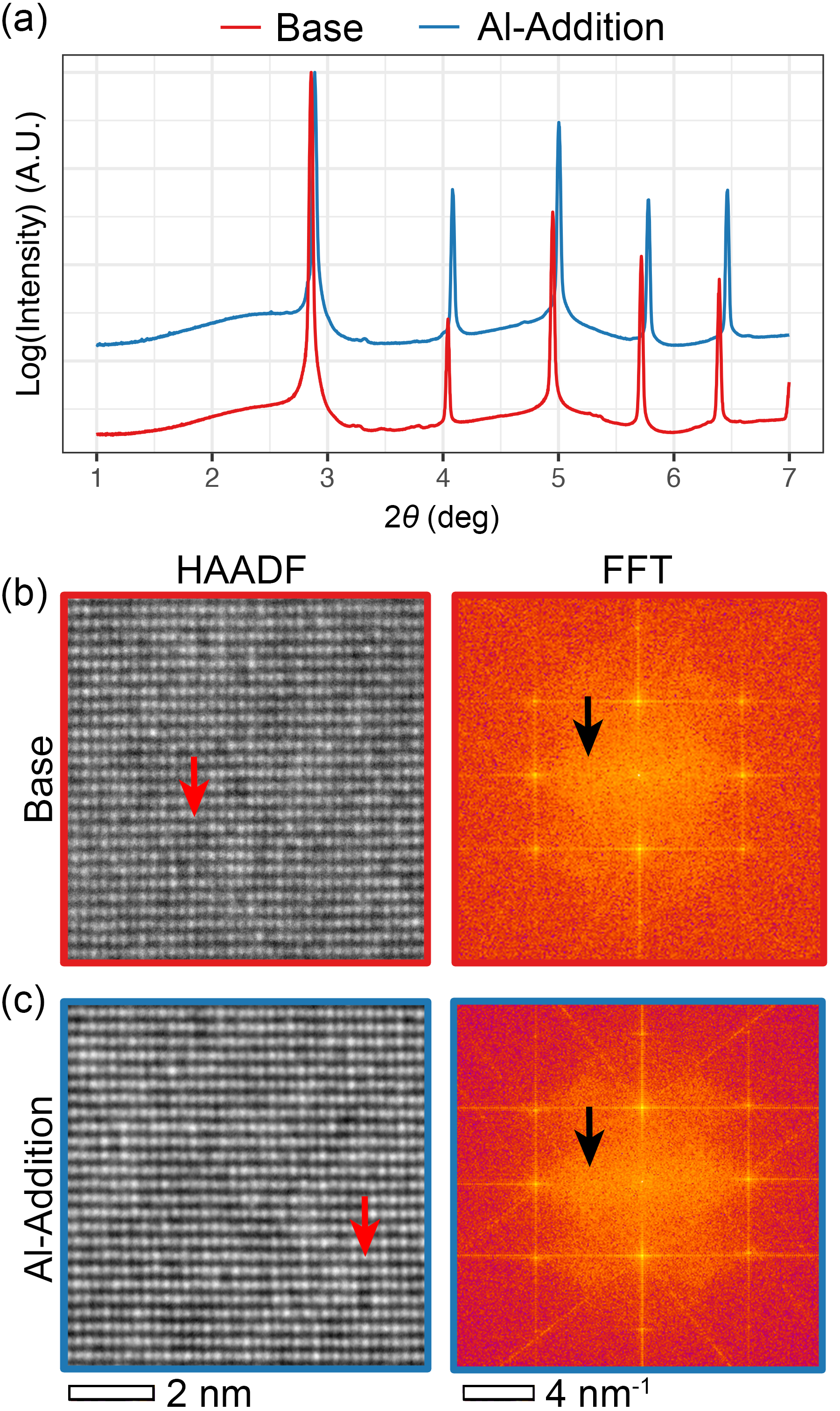}%
\caption{\label{fig:synchrxrd} (a) Synchrotron X-ray diffraction profiles for the base and Al-addition alloys.  HAADF STEM image and corresponding Fourier transform of the (b) base and (c) Al-addition alloys, showing diffuse structure. Some regions with local bright /dim atomic planes are indicated. }%
\end{figure}

Details of the base Ti$_{38}$V$_{15}$Nb$_{23}$Hf$_{24}$ and Al-addition (Ti$_{38}$V$_{15}$Nb$_{23}$Hf$_{24}$)$_{95}$Al$_{5}$ alloy fabrication and characterization  are included in Supplementary Information (also see Ref.~ \citen{Wei_naturalmixing_2020}). 
Figure~\ref{fig:synchrxrd} shows the integrated synchrotron X-ray diffraction profiles for the base and the Al-addition RHEAs. Both profiles correspond only to single phase BCC, even though the addition of Al is predicted to form distinct, long-range ordered phases. Moreover, through Rietveld refinement, Al addition is found to reduce the lattice parameter by 0.5\% (3.304 \AA{} for the Al-addition alloy compared with 3.323 \AA{} for the base alloy \cite{Wei_naturalmixing_2020}). While total X-ray or neutron scattering can reveal SRO in these systems, these techniques provide a global average of the local structure \cite{xrayabsorption, sro_heas_2021, taheri2022}. For unambiguous direct observation and comparison of SRO, especially in the case of diffuse or extremely short-range order, a more localized probe is needed \cite{niu_spindriven_2015, Zhang_Zhao_Ophus_Deng_Vachhani_Ozdol_Traylor_Bustillo_Morris_Chrzan_2019, Zhang_Zhao_Ding_Chong_Jia_Ophus_Asta_Ritchie_Minor_2020}.

Imaging both alloys at the atomic scale with HAADF STEM (Figure~\ref{fig:synchrxrd}b-c) shows strong contrast variation between neighboring atom columns, with many bright and dim bands appearing on $\{002\}$-type planes (see arrows in the figure). Based on the corresponding Fourier transforms, no discrete superlattice reflections are present, indicating the lack of long-range ordered structure. This is in agreement with synchrotron diffraction measurements. The presence of significant diffuse intensity between the $002$ Bragg spots (indicated in the FFTs in Figure~\ref{fig:synchrxrd}b-c), however, suggests that local ordering exists on $\{002\}$-type planes, or B2-type SRO, and originates from bright/dim neighboring planes in the HAADF STEM  images. 

\begin{figure}
\includegraphics{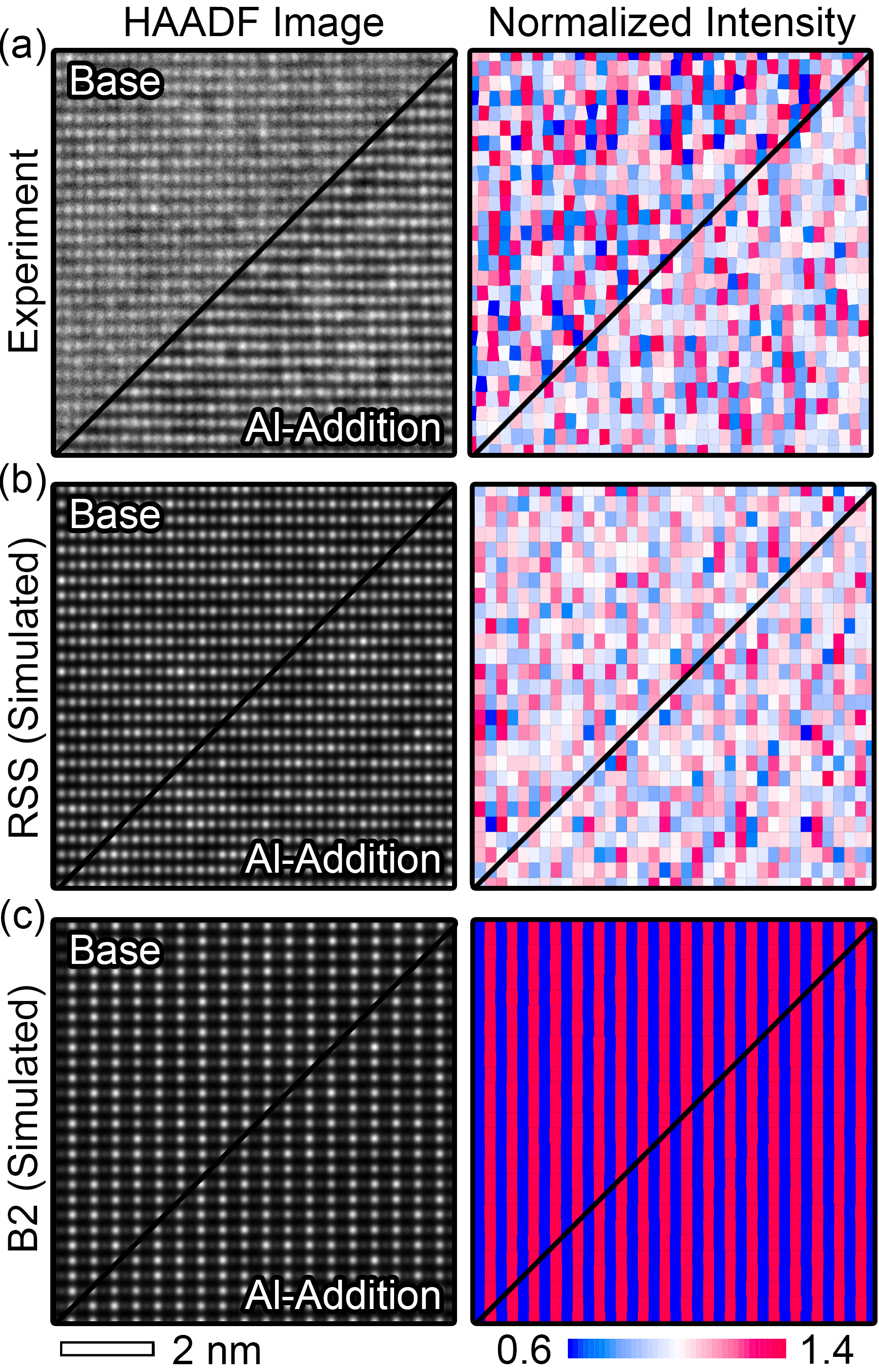}%
\caption{\label{fig:allnormalized}  HAADF STEM images and normalized intensity maps from (a) experiment, (b) ideal random solid solution STEM image simulation, and (c) ideal B2 order STEM image simulation for the base and Al-addition alloys.}%
\end{figure}

To visualize the atom column intensity variation more clearly, each atom column is normalized with respect to the average of its six nearest neighbor atom columns on adjacent $\{002\}$ planes. The resulting normalized intensity map, shown in Figure \ref{fig:allnormalized}a, highlights differences in atom column atomic number contrast \cite{niu_spindriven_2015}. This alternating intensity is present across both the base and Al-addition alloys, and at first glance suggests short-range B2-like ordering. Further, a correlated intensity (order metric \cite{kumar_pmnpt_2021}) map highlights regions of order yields hotspots of similar size and shape (Supplementary Information Figure S1). Based on the binary compound formation enthalpies (see Ref.~\citen{prxbinaryformationenthalpy}), this ordering between higher and lower atomic number elements appears to be consistent, yet the magnitudes of these enthalpies is small in comparison to other binary compound formers, including Al-containing compounds. An additional issue is apparent from comparing the normalized intensity maps of the base and Al-addition alloys. Less intense variation is seen in the Al-addition alloy, which would conventionally suggest weakened ordering \cite{niu_spindriven_2015} contrary to the effect of Al in forming ordered intermetallics and SRO in similar systems \cite{senkov2018reviewRHEASRO, prxbinaryformationenthalpy, yurchenko2022aladdition}. 

Determining the origin of the discrepancies with atomic resolution imaging is complicated by the projection-limited aspect of STEM. While hypothesis testing has been used to confidently identify SRO in systems with minimal atomic number variation, such as NiFeCrCo \cite{niu_spindriven_2015}, the wide atomic number difference between the elements in the alloys studied here introduce other issues. This is demonstrated using STEM image simulations of randomized structures with the experiment alloy compositions (Figure~\ref{fig:allnormalized}b). The results are visually very similar to the atom column contrast, normalized intensity maps, and correlated intensity (order metric) maps (see Supplementary Information Figure S1) obtained from experiment. Further, the histograms of order metric (correlated intensity, Supplementary Information Figure S2) show only slight deviation between simulated images and experiment for the Al-addition alloy.

Apparent ordering of atom column intensities is present even for the random structures based on STEM imaging. This can be explained by the presence of atoms with large atomic number differences and similar atomic percent, where the local random compositional fluctuations lead to a wide variation in contrast. Thus, there is an intrinsic amount of variation among atom column intensities purely due to the nature of STEM imaging, which introduces SRO-like contrast to these images. Depending on the composition of the alloy, differing amounts of ``pseudo-SRO'' will be present even for ideal disordered structures. This issue thus presents a significant challenge in identifying and quantifying real SRO using direct STEM imaging.



\begin{figure}
\includegraphics{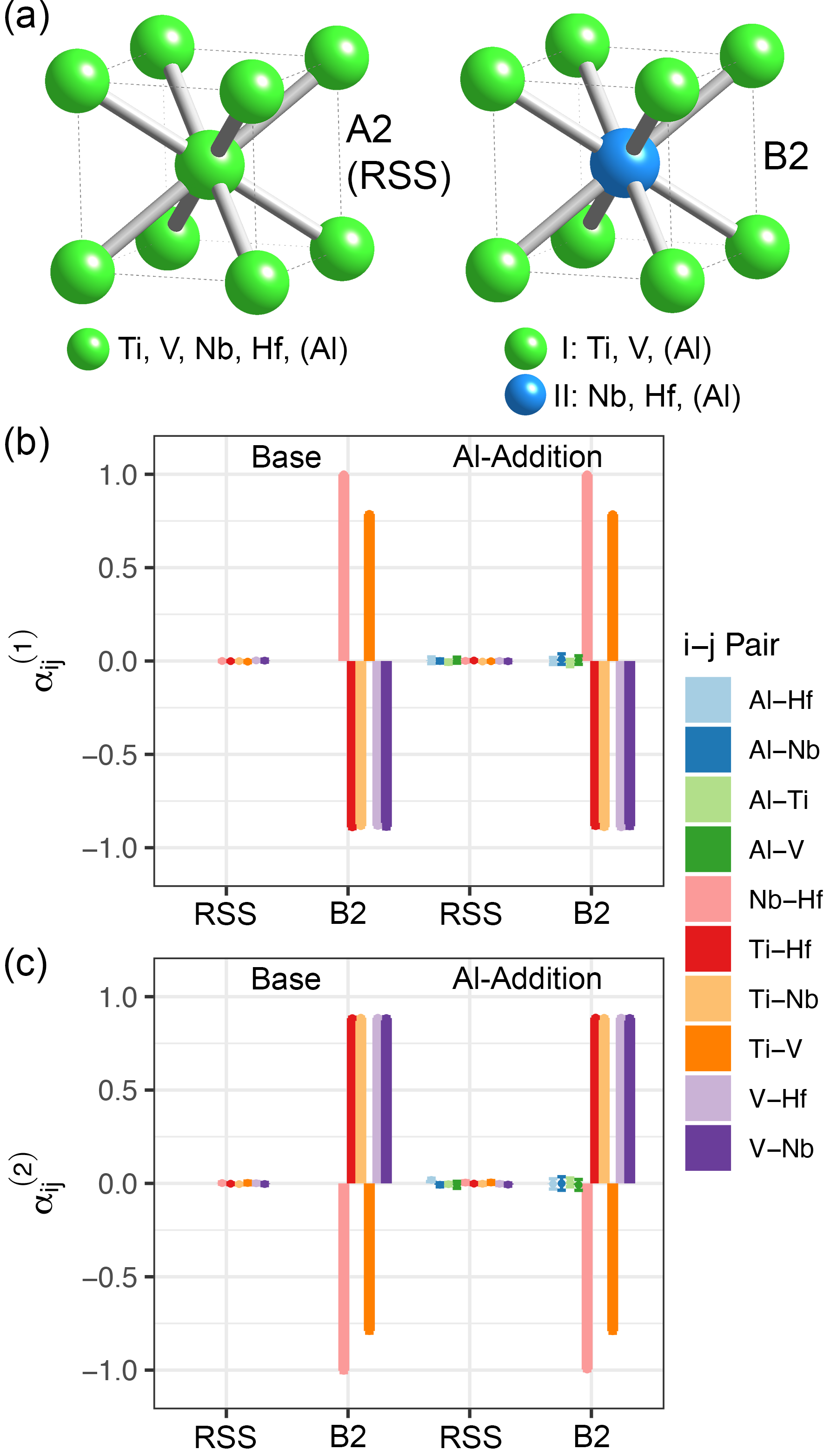}%
\caption{\label{fig:allwcs} (a) Schematic of ideal random (A2) and ordered (B2) structures with respective sublattice site occupancies. (b) First and (c) second near-neighbor Warren-Cowley pair correlations for the ideal random and ordered base and Al-addition alloy structures, averaged across seven supercells each. }%
\end{figure}

By ``bounding'' the limits of SRO using randomized and fully-ordered structures, however, statistical comparisons can be made to confidently interpret SRO from experiment and reconcile discrepancies in predicted order/disorder \cite{Xu_Kumar_LeBeau_2022}. 
Using the bulk compositions as constraints, STEM images of seven random (A2) and seven fully B2-ordered structures for both the base and Al-addition alloys are simulated using the multislice method \cite{multislice}. For the random structures, elements are randomly dispersed throughout the entire BCC $<$110$>$-oriented and thickness-matched supercells to be consistent with the bulk composition (Figure~\ref{fig:allwcs}a). In the case of fully B2-ordered structures, Al is randomly dispersed throughout the supercell, where Ti/V and Nb/Hf are randomly partitioned onto separate sublattices, with the excess Ti/V dispersed onto the other sublattice (Figure~\ref{fig:allwcs}a). 

To verify ordering in the simulated supercells, the Warren-Cowley pair correlations are used, given by
\begin{equation}
    \alpha^{mn}_{ij} = 1-\dfrac{P(_{i}^{m}|_{j}^{n})}{c_{i}} 
\label{eq:wcp}
\end{equation}
where $P(_{i}^{m}|_{j}^{n})$ is the conditional probability of finding an $i$ atom at site $m$ given a $j$ atom at site $n$, and $c_{i}$ is the overall concentration of $i$. Symmetry of the lattice reduces the independent parameters to a function of neighbor shell $k$, giving $\alpha^{k}_{ij}$, which is evaluated to the second near neighbor shell \cite{cowley1950theory, cowley1971crystals}. The resulting pair correlations confirm the ordered and disordered nature of B2 and random structures, respectively (Figure~\ref{fig:allwcs}b-c). 


\begin{figure*}
\includegraphics{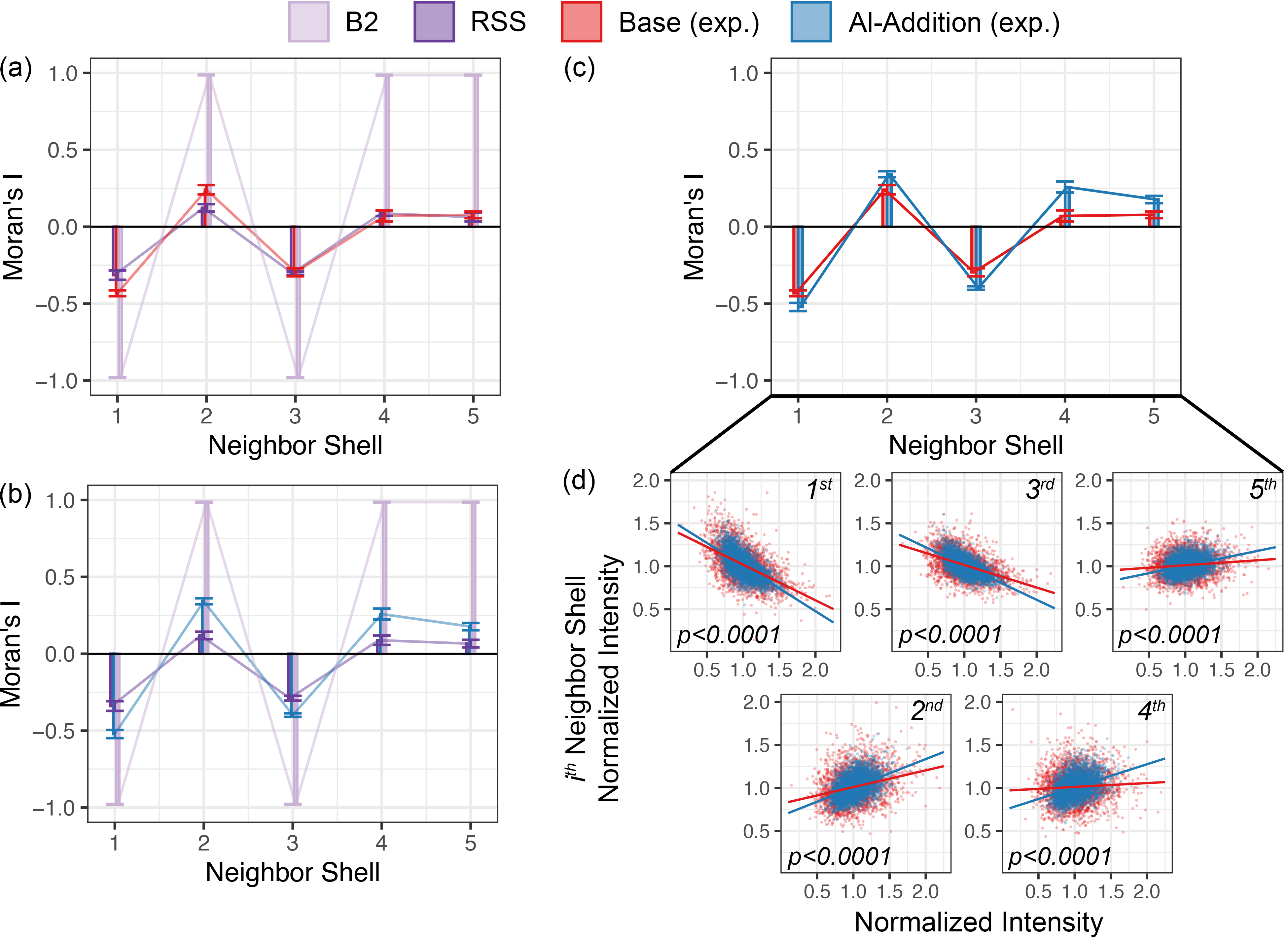}%
\caption{\label{fig:comparisononly} Bounded Moran's I values across five near-neighbor shells calculated from the normalized intensity STEM image maps for the (a) base alloy and (b) Al-addition alloy. (c) Comparison of Moran's I values between base and Al-addition alloys, with (d) Moran's I scatter plots and corresponding statistical significance at each neighbor shell.}%
\end{figure*}

To quantitatively measure order and uncover differences between simulated and experimental STEM images, a neighbor-based Moran's I statistic approach is applied, with further details provided in Ref.~\citen{Xu_Kumar_LeBeau_2022}. 
Essentially, this metric quantifies the similarity of a feature with respect to near neighbor shells and allows for the tendency and degree of order to be determined. To account for differences in the limited field-of-view and the number of sampled atom columns, sub-regions of several experimental datasets are sampled randomly to match the simulation field-of-view, and the Moran's I ordering values are calculated individually to provide the standard deviations (error bars in Figure~\ref{fig:comparisononly}). 

For simulations of the base and Al containing alloys, significant $\{002\}$ order is seen based on the pattern and magnitude of the Moran's I values across the five nearest neighbors in projection (Figure~\ref{fig:comparisononly}a-b). In the fully ordered case, the Moran's I values are near $\pm$1 and display a pattern consistent with complete and long-range B2 order. Unexpectedly, in the random case, a pattern corresponding to $\{002\}$ order is still present, confirming the local order-like features seen from the normalized intensity maps. Thus, when atoms of large atomic number differences with near equiatomic ratios are being investigated with HAADF STEM, care must be taken when making conclusions about the nature of SRO.

The simulated fully ordered and disordered structures also establish a baseline with which STEM measurements of order cannot be distinguished from  randomized site occupancy . In other words, Moran's I ordering ``greater'' than that of the random case is thus required to separate fully disordered structures from ones with true SRO. The experimental results from the base and Al-addition TiVNbHf alloys satisfy this criterion -- the Moran's I ordering values are greater than the ideal disordered structures at varying neighbor shells. Moreover, the values are between the random and fully B2 configurations. This is expected for SRO, where the random and long-range ordered structures bound the experimental Moran's I autocorrelations. The error bars (standard deviations) of the Moran's I measured across individual simulations or sub-regions are non-overlapping as well, further suggesting the presence of non-random B2 SRO. 

To confirm the statistical significance of differences in ordering between simulated structures and experiment, the Moran's I scatter plots for each neighbor shell, from which the Moran's I values are calculated, are compared using linear regression and least-squares means \cite{lstrends}. The relationship between normalized intensity and alloy composition is shown to be significant in determining near-neighbor normalized intensity, or the degree of correlated local (neighbor) order. For the base alloy, the intensity pattern found in the STEM images is distinct from random structures, with enhanced B2-like order up to the second near neighbor shell (Figure~\ref{fig:comparisononly}a), while in the Al containing alloy, ordering is statistically distinct up to the fifth near neighbor shell (Figure~\ref{fig:comparisononly}b). 

The increase in the degree and spatial extent of SRO in the Al containing alloy is verified by comparing its Moran's I values to those of the base alloy from experiment (Figure~\ref{fig:comparisononly}c). 
At each neighbor shell, the Moran's I linear fits are statistically distinct between the two compositions at a high significance threshold ($p<0.0001$), discounting the null hypothesis of no difference between the two linear regressions. Further, the larger magnitude slopes of the Moran's I fits for the Al-addition alloy indicate greater B2-like order (Figure~\ref{fig:comparisononly}d). Thus, at the local level and prior to any secondary phases, Al-addition is shown to enhance SRO formation in TiVNbHf(Al). The correlation length of order in the Al-addition alloy also increases to the fifth nearest neighbor shell in comparison with the base composition, validating the subtle increase in SRO domain radius from 0.53 nm to 0.64 nm estimated from the correlated intensity maps (Supplementary Information Figure S3). Moreover, the enhancement of SRO by Al modulation is consistent with the binary compound formation enthalpies of the given alloy components as well as Al-containing ordered phases in other refractory systems \cite{ senkov2018reviewRHEASRO, prxbinaryformationenthalpy, yurchenko2022aladdition}. 

In conclusion, local random chemical fluctuations in a near-equiatomic multicomponent alloy combined with the nature of HAADF STEM  imaging can produce image intensity variations that resemble SRO. To overcome these challenges, fully disordered and ordered supercells can be used to simulate STEM images, with resulting spatial analysis establishing bounds for the experimental results. Armed with these metrics, the results from experiment confirm the presence of local chemical order in TiVNbHf and provide further insights into enthalpy-dependent bond preferences in these systems. Further, the addition of Al is shown to enhance SRO formation at the local level, offering a route to property enhancement through microstructural control. Future work can build on these results by using the approach to provide quantitative metrics of extremely short-range order to connect to the properties of RHEAs and other compositionally complex systems.


\begin{acknowledgments}
Synchrotron X-ray diffraction measurements were carried out at beamline 11ID-C at the Argonne National Laboratory, Chicago, U.S.A., with the kind support of Dr. Andrey Yakovenko. 
We acknowledge support for this work from the National Science Foundation (CMMI-1922206). This work made use of the MIT.nano Characterization Facilities. 
\end{acknowledgments}

\section*{Data Availability Statement}
The data that support the findings of this study are available from the corresponding author upon reasonable request.


%
%

%

\bibliography{references}
\end{document}


\title[]{Supplementary Information for: Determination and Statistical Comparison of Local Short-Range Order in TiVNbHf(Al) using STEM}
\author{Michael Xu}
\author{Shaolou Wei}
\author{Cemal C. Tasan}
\author{James M. LeBeau}%
\affiliation{ 
Department of Materials Science and Engineering, Massachusetts Institute of Technology, Cambridge, MA 02139
}%


\date{\today}
\maketitle

\section{Alloy Fabrication}
The base Ti$_{38}$V$_{15}$Nb$_{23}$Hf$_{24}$ and Al-addition (Ti$_{38}$V$_{15}$Nb$_{23}$Hf$_{24}$)$_{95}$Al$_{5}$ \linebreak
(or Ti$_{36.1}$V$_{14.25}$Nb$_{21.85}$Hf$_{22.8}$Al$_{5}$) alloys employed in the current study were fabricated using raw materials (Ti, V, Nb, Hf, and Al) of $>$99.9\% purity. Vacuum arc melting was carried out using an Edmund B\"uhler GmbH AM500 arc melter with a vacuum level of 5$\times10^{-5}$ mbar. Both alloys were melted, flipped, and re-melted at least five times to avoid chemical heterogeneity before being suction cast into rectangular pieces. During arc-melting, a Zr-getter was used to avoid oxygen contamination. The as-cast ingots were cold-rolled to a 55\% thickness reduction, sealed in quartz ampoules, homogenized at 1000$^{\circ}$C for 5 hours, and then water-quenched. These specimens were next cold-rolled again to a 50\% thickness reduction, sealed in quartz ampoules, and recrystallized at 1000$^{\circ}$C for 10 min. It has been demonstrated in an earlier report \cite{Wei_naturalmixing_2020} that the amounts of interstitial C, N, and O atoms (if any) in the specimens processed using the foregoing routes are low and would not affect the current STEM measurements. 

\section{Synchrotron X-ray Diffraction}
To confirm the global single phase structure of the cast alloys, synchrotron X-ray diffraction was performed at beamline 11ID-C at the Argonne National Laboratory, Chicago, U.S.A. A high energy X-ray radiation source with a wavelength of 0.1173 Å and a beam size of 0.5 mm was adopted for the measurements. Two-dimensional diffractograms were acquired at a nominal working distance of 1600 mm and an exposure time of 0.1 s for a total duration of 10 s. These parameters were chosen on the basis of our previous work in which a suitable balance between spatial resolution and statistical representative has been confirmed \cite{Wei_Xu_LeBeau_Tasan_2021, Wei_Tasan_2020}. These two-dimensional diffractograms were post-analyzed in a GSAS-II open-access software \cite{GSAS2} to perform Rietveld refinement and integration. 

\section{Scanning Transmission Electron Microscopy}
Samples for electron microscopy were prepared from both alloys by dry mechanical polishing to under 50 $\mu$m, from which 3 mm discs were punched out and Ar-ion milled to perforation (Fischione 1051 TEM mill). High-angle annular dark-field (HAADF) images along the $<$110$>$ zone axis were acquired with a probe aberration-corrected Thermo Fisher Scientific Themis Z S/TEM operated at 200 kV using a convergence angle of 18.8 mrad and collection angle of 63-200 mrad. Image series are drift- and scan distortion-corrected using the Revolving STEM (RevSTEM) method \cite{Sang_LeBeau_2014}. Atom column intensities were extracted from the images using Python-based 2D Gaussian fitting. 

\begin{figure}
\includegraphics{figs/All Normalized_S.png}%
\caption{\label{fig:S_norms} Expanded figure from the main text: STEM HAADF images, normalized intensity maps, and correlated intensity (order metric) maps from (a) experiment, (b) ideal random solid solution STEM image simulation, and (c) ideal B2 order STEM image simulation for the base and Al-addition alloys.}%
\end{figure}

\begin{figure}
\includegraphics{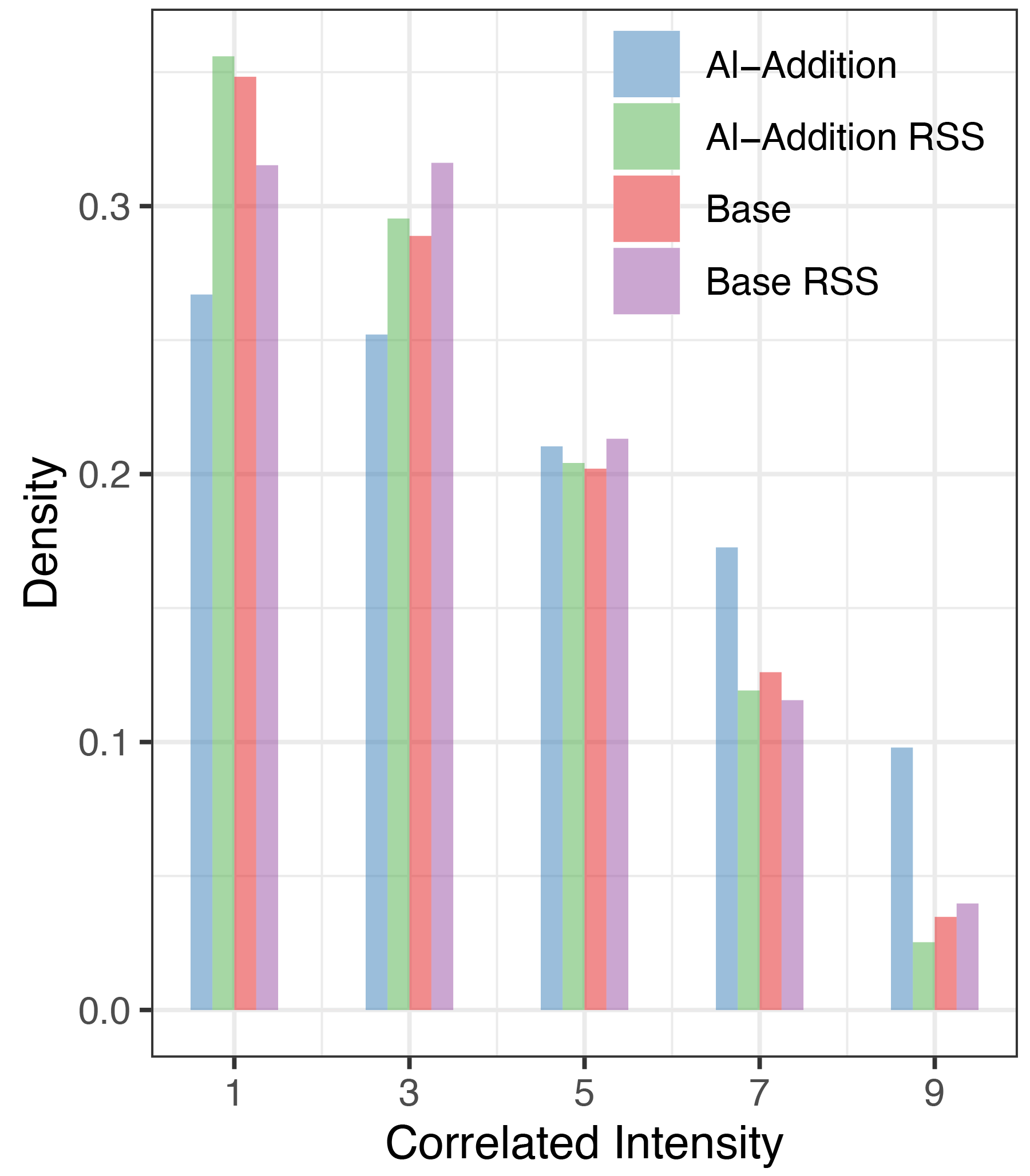}%
\caption{\label{fig:S_hist2} Distribution of order metric (correlated intensity) for base and Al-addition simulated RSS and experiment. }%
\end{figure}

Figure~\ref{fig:S_norms} shows additional correlated intensity maps, or a local order metric based on the normalized intensity maps. Correlated intensity is calculated using a correlation template in order to highlight areas of B2 order \cite{niu_spindriven_2015, Xu_Kumar_LeBeau_2022}. Similar size and shape of the ordered hotspots are present across the STEM images from experiment and simulation for both the base and Al-addition alloys. Comparing the distribution of order metric between the simulated images and experiment (Figure~\ref{fig:S_hist2}), a slight increase in amount of order (correlated intensity $\geq$7 appears to be present for the Al-addition alloy compared to RSS, but differences between other structures are ambiguous. 

\begin{figure}
\includegraphics{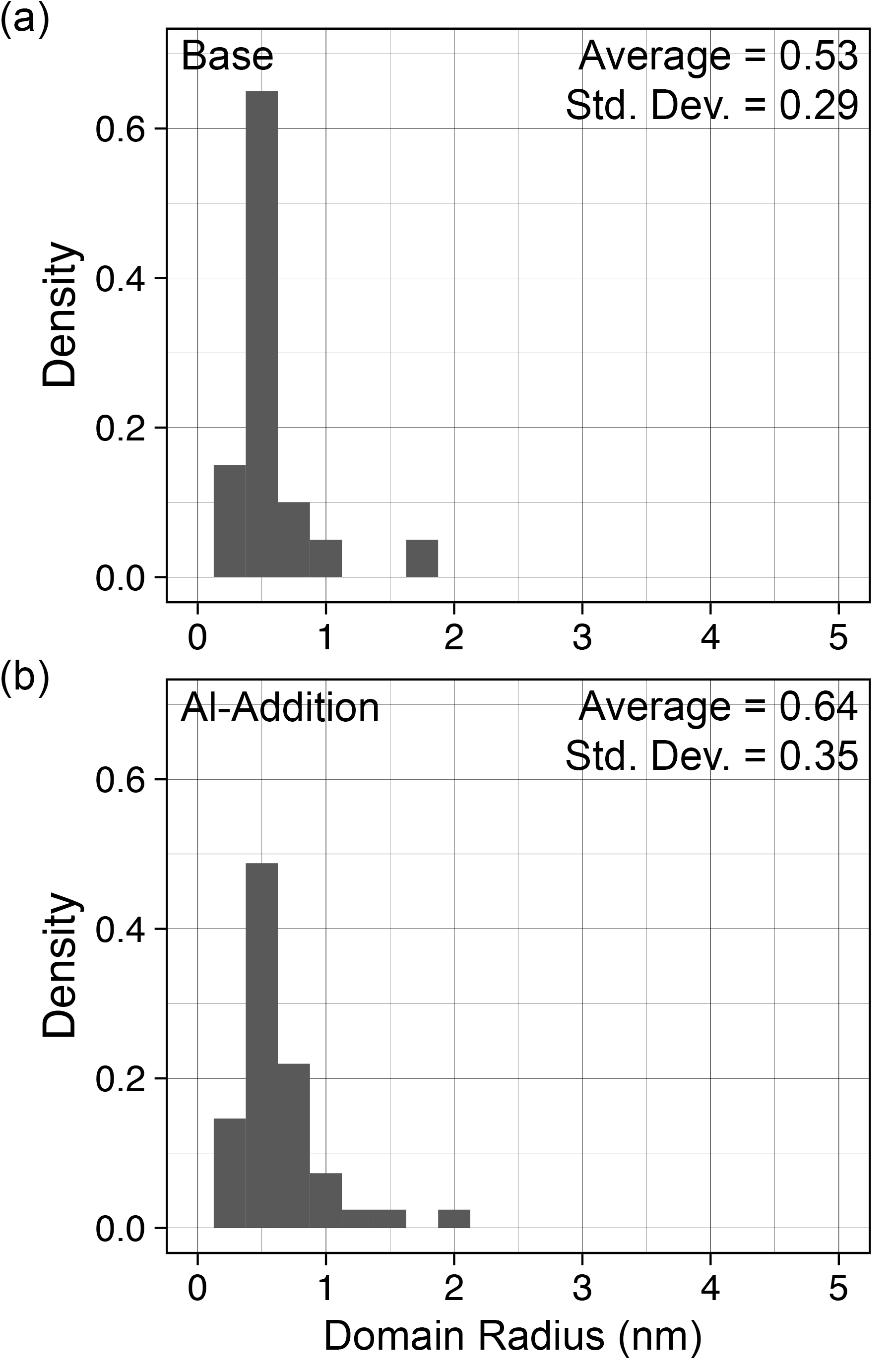}%
\caption{\label{fig:S_hist} SRO domain radius for the (a) base and (b) Al-addition alloys estimated from correlated intensity maps. }%
\end{figure}

Figure~\ref{fig:S_hist} shows the SRO domain radius estimates based on the correlated intensity maps in Figure~\ref{fig:S_norms}. Areas of high order, or order metric $\geq7$, are used to calculated domain area, from which the radius is determined. A subtle increase in the length scale of SRO is seen for the Al-addition alloy compared to the base alloy.

\bibliography{references.bib}